# Characterizing the Infection-Induced Transcriptome of *Nasonia vitripennis* Reveals a Preponderance of Taxonomically-Restricted Immune Genes

**Timothy B. Sackton[1]\*, John H. Werren[2], Andrew G. Clark[3]**

1 Department of Organismic and Evolutionary Biology, Harvard University, Cambridge, Massachusetts, United States of America, 2 Department of Biology, University of Rochester, Rochester, New York, United States of America, 3 Department of Molecular Biology and Genetics, Cornell University, Ithaca, New York, United States of America

## Abstract

The innate immune system in insects consists of a conserved core signaling network and rapidly diversifying effector and recognition components, often containing a high proportion of taxonomically-restricted genes. In the absence of functional annotation, genes encoding immune system proteins can thus be difficult to identify, as homology-based approaches generally cannot detect lineage-specific genes. Here, we use RNA-seq to compare the uninfected and infection-induced transcriptome in the parasitoid wasp *Nasonia vitripennis* to identify genes regulated by infection. We identify 183 genes significantly up-regulated by infection and 61 genes significantly down-regulated by infection. We also produce a new homology-based immune catalog in *N. vitripennis*, and show that most infection-induced genes cannot be assigned an immune function from homology alone, suggesting the potential for substantial novel immune components in less well-studied systems. Finally, we show that a high proportion of these novel induced genes are taxonomically restricted, highlighting the rapid evolution of immune gene content. The combination of functional annotation using RNA-seq and homology-based annotation provides a robust method to characterize the innate immune response across a wide variety of insects, and reveals significant novel features of the *Nasonia* immune response.

**Funding:** This work was supported in part by NIH grant R01 AI064950 to AGC and NIH grant R24 GM084917 to JHW. The funders had no role in study design, data collection and analysis, decision to publish, or preparation of the manuscript.

**Competing Interests:** The authors have declared that no competing interests exist.

\* E-mail: tsackton@oeb.harvard.edu

## Introduction

Host-pathogen coevolution is one of the major drivers of adaptive evolution across a wide variety of organisms: genes encoding proteins involved in the immune response are among the most common targets of positive selection in numerous taxa, including mammals [1,2] and insects [3–7]. There is also growing evidence that the protein-coding repertoire of the innate immune system is rapidly evolving in insects [5,8–11], even though comparative genomic analysis reveals the existence of a conserved core of immune signaling pathways in most insects [12–15]. This trend is particularly apparent among genes encoding proteins involved in recognition of pathogens and among genes encoding pathogen killing molecules such as antimicrobial peptides (AMPs), based on the few cases that have been well characterized to date. These include the taxonomically-restricted proteins Hemese [5], drosomycin [16,17], and APL1 [16,17], and lectin-24A [18], all of which are involved in recognition or removal of pathogens. Other less well characterized genes associated with the immune response, such as edin [19], Turadots [20], and IRP30 [21], also show taxonomically-restricted patterns of homology. These "taxon-specific" immunity genes likely represent both genes that evolve so rapidly that their homology to other genes is obscured and *de novo* evolution of immunity genes within taxa.

A major limitation of attempts to characterize patterns of change in the immune system across diverse insects is that, with few exceptions (*e. g.* [22]), insect immune proteins have only been well characterized on a large scale in Dipterans, primarily in *Drosophila melanogaster* [23]. While some antimicrobial peptides have been identified by proteomic approaches in diverse insect species [24–26], the bulk of genomic analyses have relied on homology to Dipteran proteins for annotating immune genes [8,9,14]. However, these methods have severe limitations for the annotation of taxonomically-restricted genes. Despite the increasing power of *ab initio* methods of gene prediction, it is still difficult to distinguish biologically meaningful predicted gene models from statistical noise in the absence of either homology information, which is by definition lacking for taxonomically-restricted genes, or functional information. Thus, despite the rapid increase in available genome sequence across insects, our functional knowledge of the diversity of immune-related genes has not kept pace.

With nine published genome sequences, including those of seven ant species [27–32], one bee species [13], and one parasitoid wasp species [15], Hymenoptera is now second only to Diptera in number of sequenced genomes among insect orders. In addition, the Hymenoptera are basal among holometabolous insects [33], making functional and comparative genomic studies in these insects relevant for the study of the evolution of immunity genes in other holometabolous taxa. In conjunction with the availability of genome sequences, several studies have used homology-based approaches (primarily with reference to *Drosophila*) to identify conserved immune-related genes in honeybees [8], wasps [15,34,35] and ants [36,37]. These studies indicate a possible reduced complement of immune-related genes in social Hymenoptera relative to Nasonia and Diptera [38]. Although unbiased studies of infection-regulated gene expression in Hymenoptera are starting to appear (e. g. [22,39]), most previous work has focused on proteomic or microarray assays that depend on existing gene annotations (e. g. [24,40,41]). This makes comprehensive tests of hypotheses regarding the role of sociality in immune gene evolution difficult, as the relevant actors may not have been identified yet. Identifying any Hymenoptera-specific immune pathways or effectors that may exist is also of key importance for understanding the biology of host-pathogen interactions that may be relevant for such challenges as colony collapse disorder [42–44].

The advent of RNA-seq technology, in conjunction with full genome sequencing, provides a method for the unbiased characterization of genes regulated by infection in almost any species that can be manipulated in the laboratory. By sequencing RNA from infected and uninfected control samples and identifying genes that are differentially regulated by infection status, it is possible for the first time to identify all genes regulated by a specific immune challenge independent of homology-based annotations. While not all genes with an immune function will necessarily be regulated by any specific infection, and thus an RNA-seq-based catalog of immune-related genes cannot be expected to be complete, it has the significant advantage of being unbiased by assumptions about conservation of genes across species, and thus serves as a valuable starting point for comparative analysis of the evolution of immune pathways.

Here, we report the infection-induced transcriptome of the non-eusocial parasitic wasp Nasonia vitripennis, combined with a new homology-based characterization of immune-related genes in the Official Gene Set V2.0 (OGSv2: http://arthropods.eugenes.org/EvidentialGene/nasonia/). Together, these form the most complete annotation of immune-related genes in N. vitripennis to date, and represent one of the first genomic-scale annotations of novel immune-induced transcripts in Hymenoptera. We show that a significant fraction of the most highly induced genes are taxonomically restricted to wasps or Hymenoptera, suggesting the possibility of previously unknown Hymenoptera-specific immune pathways. We demonstrate that a comprehensive characterization of immune-induced transcripts is feasible using transcriptome sequencing in combination with homology-based methods, paving the way for more detailed functional studies in a wider range of species and a more complete understanding of the degree to which Hymenoptera possess novel immune pathways.

## Results

### Updating the Homology-based Nasonia Immune Catalog

Several previous efforts have sought to identify Nasonia genes by homology to genes with immune functions in other species [15,34,35]. With the substantial increase in predicted gene number and gene quality obtained with the OGSv2 gene set, it is clear that an updated annotation of infection responsive genes would be valuable. We use three sources of information to update the homology-based immune catalog in Nasonia: a previously published catalog of antimicrobial peptides [35], homology to characterized Dipteran immune proteins based on reciprocal best blastp hits to D. melanogaster, and profile hidden Markov models (HMMs) of known immune-related gene families derived from ImmunoDB [11] (see methods for details).

Based on these sources, we identify 497 genes with possible immune function in Nasonia (Table S1), including 75 encoding effectors (AMPs, proteins in the prophenoloxidase cascade, transferrins, and peroxidases), 96 encoding recognition proteins (including PGRPs, TEPs, Nimrod-like proteins, and numerous lectins), 101 encoding signaling proteins, and 225 encoding proteases or protease inhibitors. We consider proteases and protease inhibitors separately because it is difficult to distinguish those with an immune role from those with other primary roles in the organism, and thus all of the immune-related proteases and protease inhibitors predicted by HMM or blast should be considered only tentatively assigned to an immune role. We discuss the other main classes in more detail below:

**Effectors.** Building upon previous work [35], we find a diversity of AMP types, including defensin-like peptides, peptides with biased amino acid composition, and helical peptides. We identify 5 defensins in two multigene families in the OGSv2 annotations, three of which are identical to those previously described [35]; we also identify 13 of the 16 defensin-like peptides (nasonins and navitricins) previously characterized [35], plus an additional possible nasonin paralog, for a total of 13 or 14 of these peptides.

In addition to the defensin-like peptides, we find all four previously characterized abaecin-like peptides and both hymenoptaecin-like peptides, all of which show a strong compositional bias toward glycine and/or proline. We also find 6 of the 12 additional Gly-rich and His-rich peptides previously characterized [35], the previously identified linear alpha-helical peptide nahelixin, and three previously identified proteins characterized by an inhibitor cysteine knot (ICK). Overall, based on homology and structural evidence alone, we find evidence for 34 AMPs in the OGSv2 gene set. We identify several other effector types as well (Table S1), including 11 members of the prophenoloxidase cascade, two lysozymes, 3 transferrins, and 18 peroxidases.

**Recognition.** We identified members of most classes of recognition proteins found in other insects (Table S1), including 3 β-glucan binding proteins, 12 PGRPs, 3 TEPs, and 9 proteins with NIM repeats. Among other putative recognition proteins, we find 34 lectins (both C-type and galectins) and 23 scavenger receptors. While it is likely that some of the lectins and scavenger receptors have non-immune functions, Nasonia appears to have a diverse recognition repertoire on par with most Dipterans and higher for at least some classes than many other Hymenopterans studied to date [8,38].

**Signaling.** As previously reported, N. vitripennis possesses all the main components of the Imd, Toll, and JAK/STAT pathways [15], and this new study does not significantly update the known signaling repertoire (Table S1). In most cases signaling proteins are both well-conserved and have a broad taxonomic distribution, suggesting that homology-based approaches will detect them easily.

We emphasize that these homology-based annotations, like previously reported annotations in Nasonia [15,34], are tentative in that they are not derived from direct functional assays. While homology-based approaches are likely to be reliable for well-conserved pathways such as the core signaling components of the innate immune network, it is difficult to confidently assign function to the numerous proteases, protease inhibitors, and similarly diverse gene families in the absence of direct functional information. While it is not straightforward to directly compare the immune repertoire reported here with previous reports that have used different underlying gene models [15,34], we do note that our inference about the number and identity of signaling components is consistent with previous annotations [15,34], while our inference about recognition and effectors tends to reflect greater, albeit still relatively minor, differences.

## Sequencing and Mapping the Nasonia Infection-regulated Transcriptome

As a complement and extension of our new homology-based immune annotation of the OGSv2 gene models, we characterized the infection-induced transcriptome in *N. vitripennis* to identify genes up-regulated by infection. To induce a broad immune response in *Nasonia* wasps, we infected 50 female adult wasps (strain AsymC) by piercing the cuticle with a needle dipped in a mixed bacterial culture containing *Serratia marcesens* and *Enterococcus faecalis*. At eight hours post infection, we separately pooled and froze the 50 infected wasps and 50 matched uninfected controls, and sequenced each pool using standard Illumina protocols to identify genes expressed in naïve and infected wasps. We mapped sequencing reads to the Nvit1.0 assembly [15] using a TopHat2 [45] and Stampy [46], and then mapped reads which failed to align to the genome against the OGSv2 predicted transcriptome with bowtie2 [47] to improve mapping sensitivity (see methods for details).

We recovered 19.77 million high quality uniquely aligned reads from the infected library, and 14.64 million high quality uniquely aligned reads for the uninfected library. We calculated counts of reads aligned to each OGS v2.0 gene model using the htseq-count script provided as part of the HTSeq python package. Most successfully mapped reads (84.0% for the uninfected library, 86.5% for the infected library) overlap a gene model, suggesting we are not missing large numbers of transcribed but unannotated genomic regions.

Using a negative binomial approach implemented in the R/Bioconductor package DESeq2 [48], we identify 183 genes significantly up-regulated by infection at a 10% FDR, and 61 genes significantly down-regulated by infection at a 10% FDR (Figure 1; Table S2). Out of 24,389 OGSv2 gene models, we did not detect any expression in either sample for 2,950 genes and filter an additional 7,289 genes because of low average expression (leaving little or no power to infer differential regulation for these genes). Of the remaining 14,150 genes which could be analyzed for differential expression, 1.29% are up-regulated by infection and 0.43% are down-regulated by infection. We discuss these infection-regulated genes in more detail in the next sections.

## Expression of Homology-annotated Immune Genes

We initially examined the patterns of regulation after infection across our homology-annotated immune classes. As expected, the 497 immune genes we describe above by homology are significantly over-represented among induced genes: 5.6% of immune genes with detectable expression are up-regulated after infection, compared to 1.2% of non-immune genes (Fisher's Exact Test, $P = 2.735 \times 10^{-11}$). The effect is largest for antimicrobial peptides and recognition genes (Figure 2). Notably, however, the vast majority of induced genes are not recognized as immune-related by homology alone: just 12% are in the homology-induced immune set. This observation suggests that the bulk of genes involved in *Nasonia* immune response remain uncharacterized, with the current study being the first identification of additional candidate genes for immunity based on functional response to infection.

Next, we used our expression data to refine our homology-based annotations with functional evidence. Among effectors, 6 of the 18 defensive-like peptides previously annotated are up-regulated upon infection (Figure 3): three of five defensin homologs, and nasonin-1, -2, and -3. The remaining nasonins are either not regulated or filtered due to low expression. These peptides may have diverse functions in *Nasonia* (including non-immune functions), or respond to other pathogens not included in our infection cocktail. Beyond the defensin-like peptides, 4 of the 6 abaecin or hymenoptaecin homologs in *Nasonia* are significantly up-regulated after infection. These glycine and proline rich peptides are some of the most highly induced genes in our entire study, with hymenoptaecin-2 induced over 50-fold and abaecin-2 induced over 30-fold (Table S1). However, none of the other Gly- or His-rich peptides, the previously identified linear alpha-helical peptide nahelixin, or the three previously identified proteins characterized by an inhibitor cysteine knot (ICK) are significantly induced in our experiment. These may either represent AMP-like proteins with non-immune functions, or be induced by different pathogenic challenges.

Genes encoding recognition and signaling proteins are much less likely to be induced than genes encoding effectors (Figure 2). Nonetheless, 3 of the 12 PGRPs in *N. vitripennis* are induced by infection, along with 2 lectins (Table S1). Notably, several PGRPs are among the most-strongly induced genes after infection in *D. melanogaster* [49].

While none of the genes encoding signaling proteins have evidence for differential expression after infection, this functional class is notable for the consistency with which we can detect expression in both uninfected and infected samples. Overall, we can only detect measurable expression for 58% of OGSv2 gene models, but we detect measurable expression for all 101 genes encoding signaling proteins.

## Previously Uncharacterized, Highly-induced Genes Encode Short, Secreted Proteins

In order to assess whether previously uncharacterized but highly-induced genes represent novel immune effector molecules, we assessed which and how many of these genes encode proteins that possess properties common to known AMPs. To start, we attempted to infer whether they are secreted proteins by computing the likelihood of a signal peptide for all predicted protein sequences derived from OGSv2 gene models (see methods). Most well-characterized effector molecules in other insects are secreted into the hemolymph: in our study, 91.2% of previously identified AMPs have bioinformatic evidence indicating presence of a signal peptide. Excluding genes with homology evidence for immune function, 42.24% of induced genes encode proteins with bioinformatic evidence for a signal peptide, a significantly higher fraction that the 15.95% of non-induced and non-immune genes that encode a protein with bioinformatic evidence for a signal peptide ($P = 3.4 \times 10^{-15}$, Fisher's Exact Test).

In general and in our *Nasonia* data, antimicrobial effectors are shorter than 300 amino acids (33 of 34 overall in *Nasonia*). While induced genes with no homology evidence for immune function do not have quite as extreme a skew in encoded protein size, there is still a significantly higher proportion of genes encoding proteins less than 300 amino acids (57.1% compared to 42% in the non-induced set, $P = 1.46 \times 10^{-4}$, Fisher's Exact Test). Finally, genes encoding antimicrobial peptides (and indeed many non-signaling components of the immune system) are often members of multi-gene families present across insects [5]. In *Nasonia*, 38.5% of genes encoding AMPs are in multi-gene families, compared to 17.1% of non-immune genes (Fisher's Exact Test $P = 0.0081$). Even after excluding homology-annotated immune genes, the proportion of multi-copy genes in the induced class is significantly elevated (44.7% compared to 18.9% for non-induced genes, Fisher's Exact Test $P = 7.08 \times 10^{-13}$).

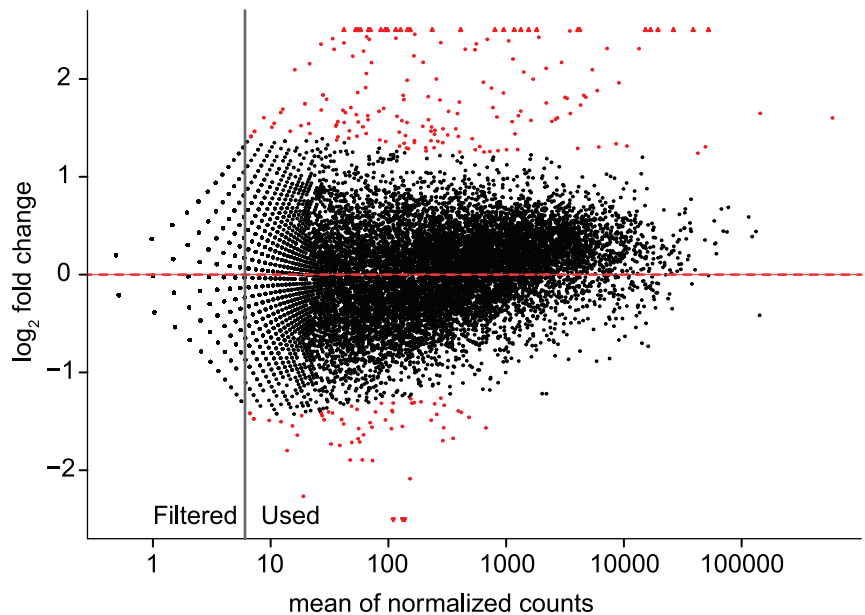

**Figure 1. MA plot of differentially expressed genes between Infected and Uninfected treatments.** Points in red indicate significance at an FDR of 10%. Points with a mean counts below the dashed line were filtered prior to multiple test correction, as genes with low mean counts across conditions have little or no power to detect differential expression (see methods for more details). Triangles represent points with log2 fold change greater than 2.5 or less than −2.5.
doi:10.1371/journal.pone.0083984.g001

## Bioinformatic Characterization of Putative Novel Effectors in *Nasonia*

Overall, 86.3% (139/161) of induced genes that have no homology-based evidence for an immune function are either members of multi-gene families, encode proteins shorter than 300 amino acids, encode proteins with bioinformatic evidence for a signal peptide, or have a combination of these properties. Although antimicrobial peptides are not the only functional class to be overrepresented among short, secreted proteins encoded by members of multi-gene families, the prevalence of these properties among induced genes strongly suggests the possibly that at least some of these genes encode novel immune effectors. We selected as candidates for more detailed characterization the 37 genes that encoded proteins that were both less than 300 amino acids and had a signal peptide. Of these, 11 have blastp hits to various diverse *D. melanogaster* proteins with no obvious immune function, although these genes may have acquired an immune function in *Nasonia* or lost an immune function in *D. melanogaster* since those two species last shared a common ancestor.

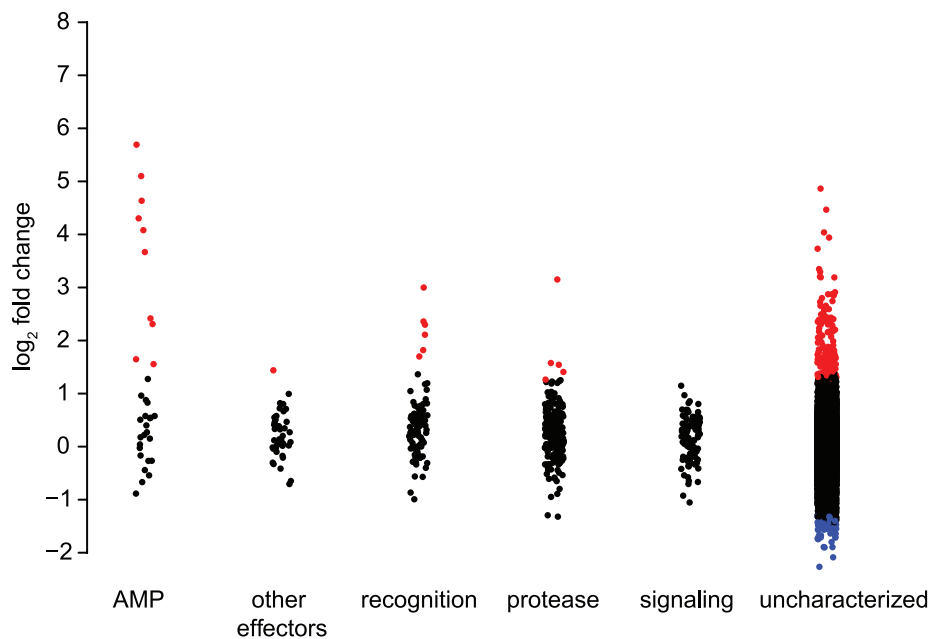

**Figure 2. Fold change of immune-annotated and non-immune annotated genes after infection.** Scatterplot (with jitter of overlapping points) showing fold-change for AMPs, other effectors, recognition, signaling, proteases, and non-immune genes. Points in red represent significant up-regulation after infection, points in blue represent significant down-regulation.
doi:10.1371/journal.pone.0083984.g002

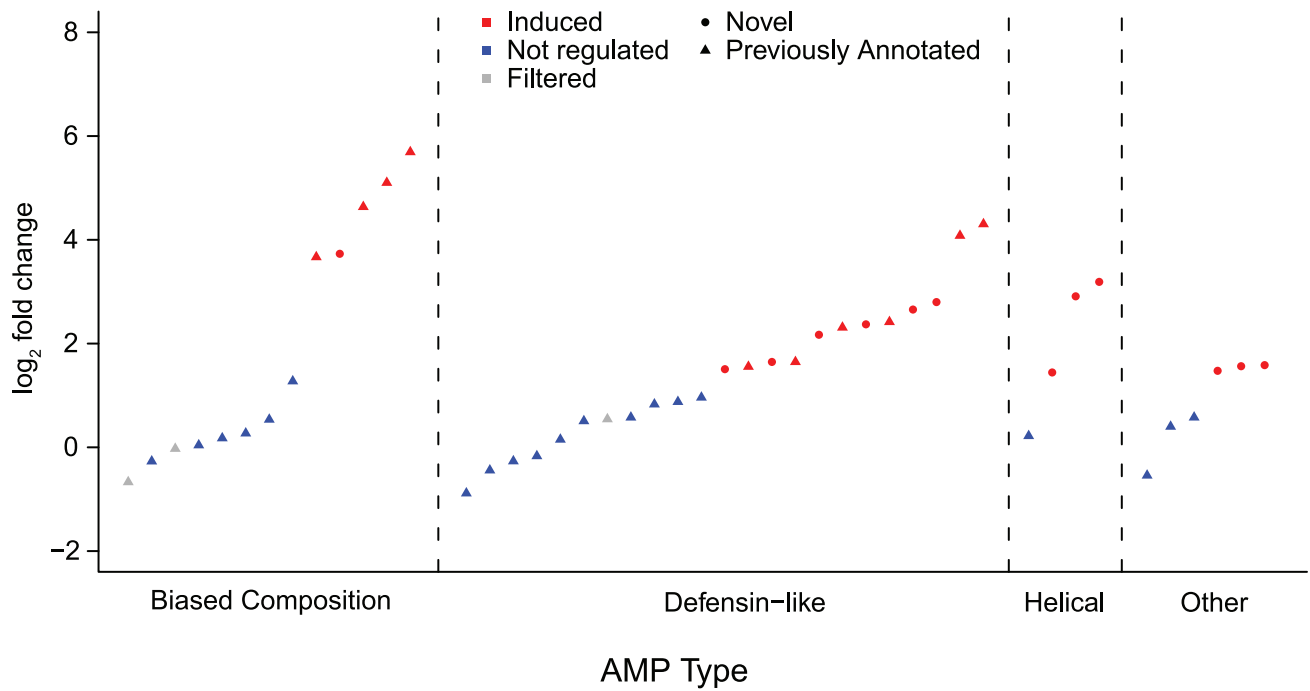

**Figure 3. Induction of known and putative novel AMPs after infection.** All 34 previously identified AMPs (excluding Nasonin-7 which does not have detectable expression in our dataset), plus the 14 putative novel AMPs, are plotted on the X-axis, ordered by AMP class and degree of induction (log2 fold change, on the Y-axis). Points are colored by differential expression class, with different shapes for novel and homology-annotated genes.
doi:10.1371/journal.pone.0083984.g003

We focused on the 26 genes encoding proteins with no blastp hit in *D. melanogaster*. Of these, 14 encode proteins with a positive net charge, suggesting the potential for an antimicrobial peptide role; their pattern of induction after infection is shown in Figure 3. Broadly speaking, these 14 proteins fall into several classes (Table S3): one extremely cysteine-rich protein that does not form obvious secondary structures; six putative defensin-like proteins characterized by beta sheets and disulfide bridges; three proteins with extensive helical domains, but likely too large to be linear alpha-helical peptides; and four other proteins with mixed secondary structures. Even though we cannot definitively assign these genes an immune effector role, it is apparent that defensin-like peptides have an unusual diversity in *Nasonia*, representing 18/34 (52.9%) homology-annotated AMPs and 6/14 (42.9%) of putative novel AMPs.

## Previously Uncharacterized, Highly Induced Immune Genes are Biased toward being Taxonomically-restricted

To better understand the evolutionary history of immune genes in *Nasonia*, we estimated gene age for all OGSv2 gene models. Briefly, we define the phylogenetic age of a gene as the deepest node for which an homologous protein can be found by blastp, without regard to gapiness of the homology pattern. For example, a *Nasonia* gene encoding a protein with a significant blastp hit to a mouse protein is assigned to the "Metazoan" stratum regardless of the presence of homologs in other species. This approach is conservative with regard to taxonomically-restricted genes, in that we are much more likely to incorrectly assign a gene to a deep stratum than incorrectly assign a gene to a recent stratum (see methods for details). We classify each gene into one of five phylogenetic strata (ages): genes that are conserved across metazoans ("Metazoan"), genes that originated before the divergence of crustaceans and arachnids from insects ("Arthropod"), genes that originated prior to the diversification of insects but after their divergence from crustaceans ("Insect"), genes that are Hymenoptera-specific ("Hymenoptera"), and genes with no homologs outside of *N. vitripennis* ("Wasp"). Because of the uneven phylogenetic sampling of sequenced Hymenoptera genomes, we cannot distinguish *Nasonia*-specific genes from genes that are common to many wasps but arose after the split from the well-sampled bee and ant lineages.

Based on this classification of evolutionary age of each gene, induced genes are much younger than either non-regulated or repressed genes: excluding those genes without significant expression, induced genes are only 37.9% Metazoan compared to 64.5% Metazoan for non-regulated genes and 55.9% Metazoan for repressed genes (Figure 4A). This difference in proportion Metazoan is highly significant ($P = 6.91 \times 10^{-13}$, Fisher's Exact Test); induced genes are also much more likely to be wasp-specific (31.3% induced, 16.3% other genes, Fisher's Exact Test $P = 8.85 \times 10^{-7}$).

A potential confounding factor, however, is gene length, as it is more difficult to detect homology among shorter genes. Rapidly diverging short genes, even when they are evolutionarily orthologous to genes in other species, may be nearly impossible to classify as such. We use two approaches to control for gene length. First, we consider only genes encoding proteins with sizes in the interquartile range of the induced class (134 to 456 amino acids, inclusive). Even restricting our analysis to genes encoding these small proteins, induced genes remain much younger than other classes (38.9% Metazoan compared to 63.9% for non-induced genes, $P = 2.24 \times 10^{-6}$, Fisher Exact Test; Figure 4B). In this restricted set, induced genes are also more likely to be wasp-specific (26.7% induced, 14.8% others, Fisher's Exact Test $P = 0.004$).

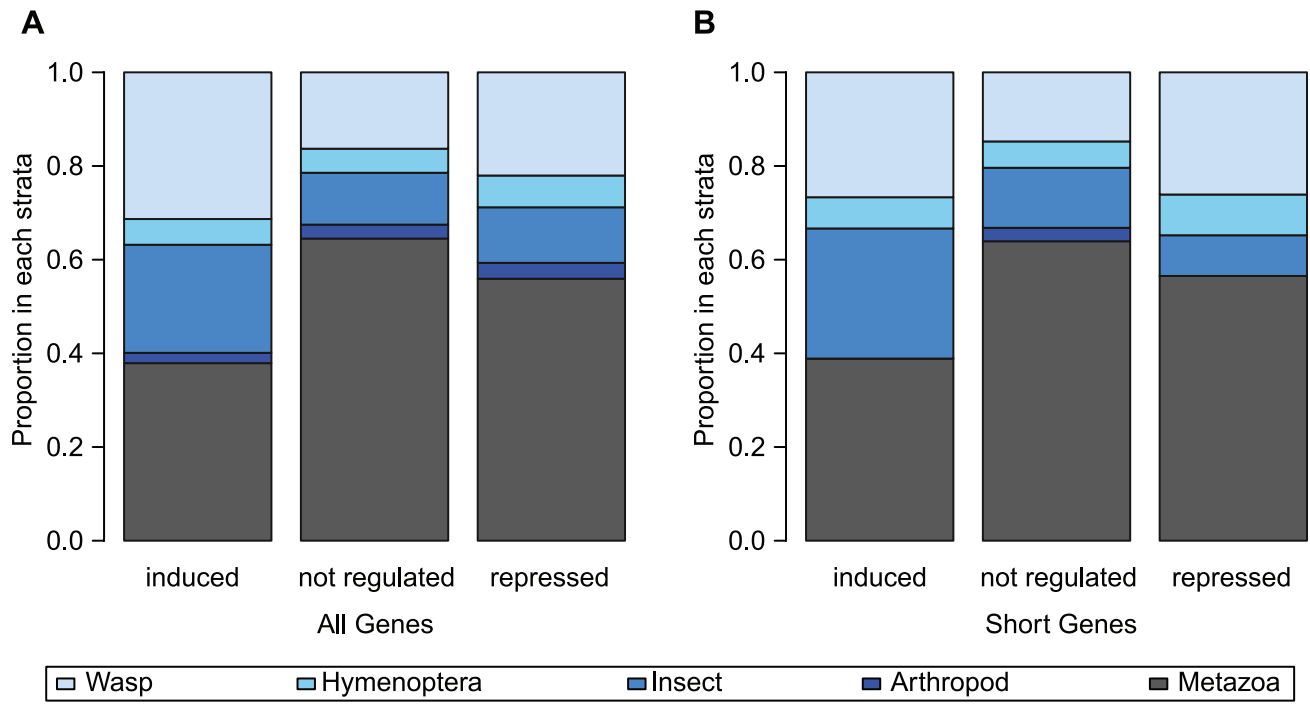

**Figure 4. Phylogenetic strata for genes of different expression classes.** A) All genes, which includes all expressed genes. B) Short genes, which includes only genes encoding proteins between 134 and 456 amino acids, inclusive.
doi:10.1371/journal.pone.0083984.g004

As a second approach, we fit a logistic regression using an indicator variable set to 0 if gene age is "Metazoa" (and 1 otherwise) as the response variable, and using gene size and an indicator variable set to 1 if differential expression class is induced (and 0 otherwise) as predictor variables. As expected, size is significantly ($P < 2 \times 10^{-16}$) negatively associated with gene youth, with larger genes less likely to be classified as young. Notably, being induced is positively associated with gene youth ($P = 2.96 \times 10^{-9}$), even when including size as a cofactor in the model. We get identical results if we define "gene youth" as being wasp-specific, instead of as not being present in the common ancestor of Metazoans. Both these lines of evidence suggest that induced genes are significantly younger than non-regulated, and repressed genes, and specifically that induced genes are more likely to be taxonomically-restricted to a subset of lineages within Metazoans.

## Discussion

The number of sequenced insect genomes is likely to grow dramatically in the next few years, as the i5K project and related efforts begin to bear fruit. The challenge now is to make biological sense of the genomic data being produced. Taxonomically-restricted genes, especially those that seem to have arisen *de novo* from non-coding sequence [50–55], are likely more crucial to organismal fitness than previously recognized [56], and it is becoming clear that these genes may be a common feature of many genomes [38,57,58]. In light of this, homology-based methods of functional gene annotation will be suitable for only a subset of the genome. While functional analysis of single genes and pathways will always have an important role to play in understanding the biology of diverse insects, such techniques are not scalable to multiple genomes across many species. RNA-seq technology offers a rapid and scalable approach to characterize how gene expression changes in response to experimental manipulations, and is applicable across a wide range of taxa.

The innate immune system is particularly amenable to characterization with expression-based methodology such as RNA-seq, as one of the key biological consequences of pathogenic infections is the rapid induction of several classes of effector proteins, along with up-regulation of a number of other pathway components. This process has been extremely well-studied in *D. melanogaster* [49,59–64], but historically has been challenging to study in other insects where molecular tools to measure gene expression have not been as well developed.

In this study, we have characterized the immune-inducible transcriptome in *N. vitripennis* using RNA-sequencing, and identified genes that respond transcriptionally to infection. Combined with an updated homology-based annotation using a profile HMM approach pioneered in mosquitoes [11], the analysis provides a wealth of functional information about the immune response in *Nasonia* wasps. Strikingly, the vast majority of induced transcripts cannot be associated with an immune function based on homology alone. While further study will be necessary to better understand the biological role these immune-inducible genes play, it is clear from this study that a transcriptional approach reveals a depth and complexity to the immune response in *Nasonia* that is not apparent from only comparative and computational annotations.

Many of the newly characterized immune-inducible genes we describe are taxonomically-restricted to either the wasp lineage, or to Hymenoptera as a whole. In some cases these seem to represent novel, previously uncharacterized antimicrobial peptides, but many more of these immune-inducible genes are difficult to characterize in the absence of detailed functional studies. For example, it is not yet clear whether these genes are regulated by the same conserved signaling pathways present in Dipterans, or whether novel immune signaling pathways may also exist in Hymenoptera. In either case, the results make clear that the immune system is a dynamic network with substantial turnover in membership over evolutionary time, as has been previously suggested [5,9–11].

The *N. vitripennis* immune response is of particular interest given the complex biology of bacterial associations documented in this species. A recent study indicates that disruption of species-specific regulation of different gut bacteria plays a role in hybrid lethality among *Nasonia* species [65], and the endosymbiotic bacterium *Wolbachia* [66] is known to cause reproductive incompatibilities among species [67,68]. Wasp genotype influences both *Wolbachia* density in host tissues [69] and the type and level of reproductive alterations induced by *Wolbachia* [70–72], implying a role of variation in host response in the biology of *Wolbachia* in *Nasonia*. While the role of the wasp immune response specifically in these interactions is not yet clear, the set of genes we identify here will provide crucial context for understanding the wasp response to these specialized bacterial interactions and allow incisive tests of the degree of overlap between the *Nasonia* response to *Wolbachia* and the more general immune response.

Within Hymenoptera, considerable interest has focused on the role that eusociality plays in shaping the evolution of the immune system [7,8,36,73], with at least some suggestion that social species may have a reduced complement of immune-related genes. However, in most cases these conclusions have been derived primarily from homology-based annotations that are ultimately based on molecular studies in *Drosophila* and mosquitoes. As we show here, it is certainly feasible that extensive Hymenoptera-specific immune components exist, and until similar studies are performed across diverse Hymenoptera species it will be difficult to fully disentangle the role of eusociality in the evolution of Hymenopteran immunity.

The current explosion of genomic sequence makes this an exciting time to study insect immunity, and although further work will be necessary to extend the analyses we present here, the combination of genome and transcriptome sequencing provides a scalable approach to characterize the inducible immune response across a broad taxonomic range.

## Materials and Methods

### Nasonia Materials, Infections, and Sequencing

Individual female adult wasps ($n = 50$) of the sequenced strain (AsymC) were infected by poking with a 0.1 mm dissecting pin dipped in a bacterial culture containing an equal mixture (as measured by OD600) of *Serratia marsecens* and *Enterococcus faecalis*. Pools of wasps from infected ($n = 50$) and uninfected ($n = 50$) treatments were frozen separately in liquid nitrogen 8 hours after experimental manipulation. Total RNA was extracted from each pool using standard Trizol protocols. Samples were poly-A selected and prepared for sequencing on an Illumina GAII following standard protocols. The infected sample was sequenced on three lanes, and the uninfected sample was sequenced on two lanes, producing a total of 53,464,989 reads from the infected sample and 16,782,852 reads from the uninfected sample. The larger number of reads per sequencing run from the infected library likely reflects the timing of the reads, as two of the three infected runs occurred after an upgrade in the capacity of the core facility. All individual lanes from the same library (infected or uninfected) were pooled prior to analysis. Raw sequencing reads are available from NCBI with accession number SRP029983, and aligned BAM files are available from NasoniaBase (http://hymenopteragenome.org/nasonia/).

### Mapping Reads to the Nasonia Genome

We aligned our RNA-seq reads to the *N. vitripennis* reference genome version 1.0 [15], available from NasoniaBase. We used a combination of TopHat2 [45] and Stampy [46] to map our sequencing reads to the reference scaffolds, in order to leverage the unique strengths of each program and maximize mapping success. We first mapped each library (Inf or Unf) using Tophat2 with the following options: -N 3–read-edit-dist 5–read-realign-edit-dist 2 -i 50 -I 5000–max-coverage-intron 5000 -M, and using the Nvit OGSv2 GFF file (http://arthropods.eugenes.org/EvidentialGene/nasonia/) as a genome reference (−G). Next, we realigned the unmapped reads from TopHat with Stampy with the following options: –substitutionrate = 0.01–sensitive. The goal of the dual mapping protocol is that while TopHat deals well with spliced reads, it is not as sensitive as Stampy in mapping reads with polymorphisms. Stampy, however, does not deal well with junctions, and is very slow. Thus, Stampy is only used to map reads that TopHat cannot accurately place. After mapping, we merged the Stampy and TopHat output to produce a single BAM file for each library. To check for 3′ bias, we estimated read coverage across the length of the gene body based on computing reads (mapped to the genome using TopHat2 or Stampy) mapping to each percentile bin of gene length using RSeQC [74]. In general we see little evidence for 3′ bias: mapped reads that overlap genes are generally relatively unbiased with respect to gene position (Figure 5).

After this mapping procedure, we were still left with a substantial number of unmapped reads, especially in the infected sample. To capture additional spliced reads that remained unmapped, we remapped all unmapped reads against the OGSv2 predicted transcriptome using the –very-sensitive-local option in bowtie2 [47], and added the counts of reads mapped to each gene to the counts derived from the genome mapping.

### Detecting Differentially Expressed Genes

To identify genes that are regulated by infection, we used the DESeq2 package [48] in Bioconductor/R. We first counted the number of reads that overlap each transcript in the OGSv2 dataset (which includes 24,389 gene models) with the htseq-count script included as part of the HTSeq Python package (http://www-huber.embl.de/users/anders/HTSeq/doc/overview.html), using the following options: -a 30 -m -intersection-nonempty -s no -t exon -i gene_name. We then processed the resulting count files with a custom perl script to merge counts across manually annotated genes that are split across scaffolds ('splitgenes' in the GFF file) and add counts from our final transcriptome mapping round. We used the resulting count file as input to DESeq2.

DESeq2 uses a negative binomial distribution for statistical inference about differential expression. Because our biological replicates were pooled prior to RNA extraction and sequencing, we ran DESeq2 with several modifications to the default procedure. We use the standard dispersion estimation approach, except that we treat the infected and uninfected samples as replicates in order to estimate dispersion. We use the standard maximum *a posteriori* approach to fit genewise dispersions, but we do not use outlier detection, nor do we use filtering on Cook's distance. We use the independent filtering approach to filter the final results on mean normalized counts across all genes, as genes with very low counts have little or no power to detect differential

**A**                                                                 **B**

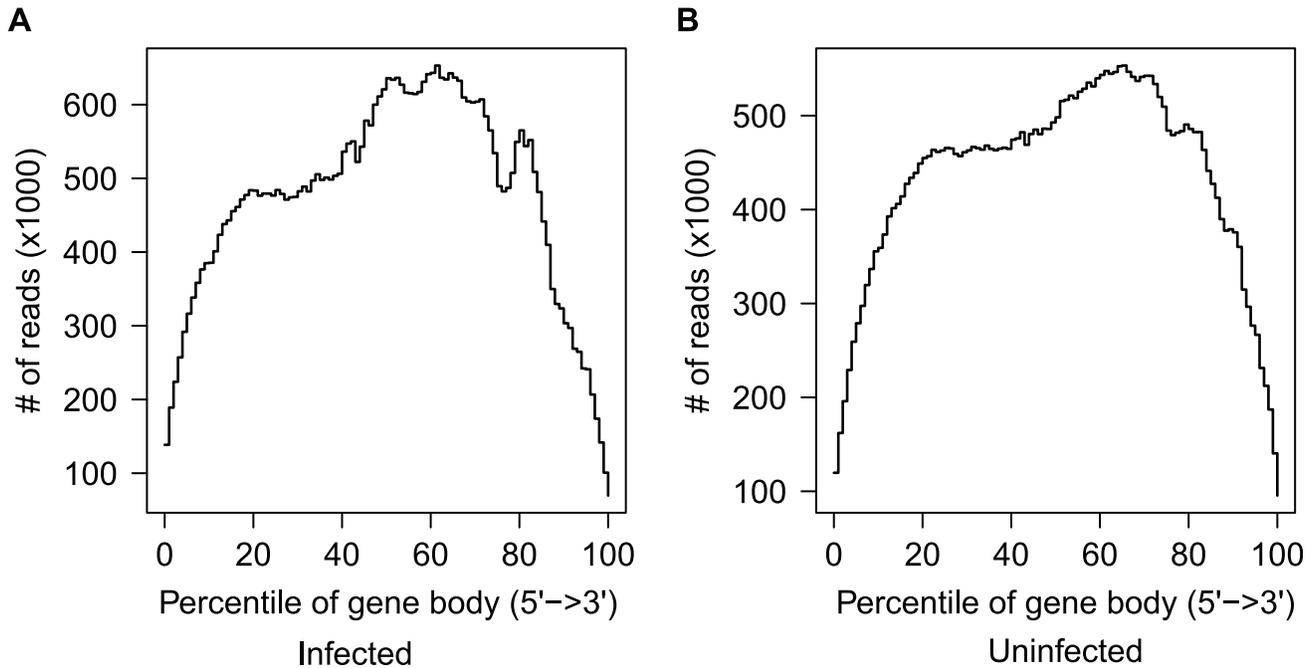

**Figure 5. Distribution of read coverage along genes for the A) infected and B) uninfected library.** For each library, the distribution of reads mapped across the normalized gene body length, as calculated by RSeQC based on reads mapped to Nvit1.0 scaffolds. Each gene is divided into bins of 0.01×length and reads mapped to each bin are summed across genes.
doi:10.1371/journal.pone.0083984.g005

expression, although we capped independent filtering cutoff to the median of the mean of normalized counts across samples (6.81).

The biological replication in our study is in the form of many individual wasps that were pooled prior to sequencing, which means we have only modest power to detect differential expression. This primarily arises because in order to be able to estimate the fit of the negative binomial distribution to our count data, we estimate dispersion by treating the infected and uninfected sample as replicates of each other and then relying on the observed mean-variance relationship. We cannot estimate among-individual, within treatment variation. However, given the large number of individuals that contributed to each RNA pool, it is unlikely that the genes for which we do detect differential expression represent false positives arising from high among-individual variance.

### Annotating the OGSv2 Gene Set

To better understand the properties of immune-regulated genes in *Nasonia vitripennis*, we merged annotations from both novel and existing annotations to infer patterns of homology and identify previous immune characterizations. The full set of annotations, including differential expression statistics, is available as Dataset S1.

**Homology-based immune annotation.** We relied on three sources to identify *Nasonia* genes with immune annotations based on homology to known immune genes. First, we add all antimicrobial peptides identified by a previous computational screen [35], filtered to remove cases where gene models were removed during a subsequent annotation or NCBI gene models could not be associated with OGS 2.0 gene models. We assign the function "antimicrobial peptide" to all these genes. Second, we extracted reciprocal best blastp hits (RBH) between *N. vitripennis* and *D. melanogaster*. For each *Nasonia* gene with a RBH ortholog in *D. melanogaster*, we assigned the immune function annotated for *D.*

*melanogaster*, based on previously published lists of immune genes in *Drosophila* [5].

Finally, we identify additional immune genes based on homology to manually curated immune gene families from ImmunoDB [11]. Here, we extracted protein sequences (either whole proteins or single domains) from the manually curated Dipteran immune gene sets available on ImmunoDB (http://cegg.unige.ch/Insecta/immunodb), aligned them with FSA [75]) (options: –fast), and generated profile HMMs with HMMER 3.0 (http://hmmer.janelia.org/). We also include an alignment of the NIM domain from Nimrod-like protein [76], which is missing from ImmunoDB. We then searched each of these 25 immune-related profile-HMMs, against all *Nasonia* proteins using hmmsearch. We corrected e-values to reflect the fact that we did 25 separate searches, and then retained all hits with corrected E-values <= 1. In the few cases where multiple searches hit the same protein, we retained the hit with the smallest E-value.

Combining all three sources of evidence, we identified a total of 497 immune genes based on homology, including 32 from previous computational screens for antimicrobial peptides, 106 based on *D. melanogaster* orthologs, and 361 based on profile HMMs. The full list, including expression and functional assignments, is available at Table S1. We note that this is a computationally-derived and homology-based list, and suffers from the limitations inherent in all such efforts.

**Orthology and paralogy.** We infer two key properties for each gene related to patterns of homology: first, the age of each gene based on the phylogenetic stratum in which each gene can be inferred to have originated; and second, the number of paralogous copies of each gene in the *Nasonia* genome. In both cases, the results are based on the same set of blastp searches, in which we blasted each *Nasonia* gene against the predicted proteomes of 30 additional species, including 12 additional Hymenoptera genomes, 6 Dipteran genomes, 5 additional insect genomes, 2 additional non-insect arthopod genomes, and 5 non-arthropod outgroup genomes (see Table S4 for the full list, including gene model versions and data sources). For the blastp search, we first softmasked the blastdb using SEG, and then used standard blastp with an e-value cutoff of 0.001. For cases where there are multiple isoforms of a gene, we include all of them but then collapse hits across isoforms, preserving the hits with the lowest e-values.

We define phylogenetic strata as the deepest node in the branch at which at least one detectable homolog is present, where a detectable homolog is defined as a blastp hit where the alignment covers at least 50% of the shorter protein with at least 30% positives. This measure allows for gapiness in the pattern of homology, so a *Nasonia* protein with a hit to a single non-arthropod genome would be defined as "Metazoan" even if it were absent from all arthropod genomes other than *Nasonia*. This biases our results towards assuming deeper origins of genes than may be the case (if, for example, some phylogenetic incongruence arises from horizontal gene transfer), but we consider this property desirable as it makes our analyses conservative. We also define paralogs as pairs of *Nasonia* genes with reciprocal blastp hits with e-values $<= 0.0001$, at least 30% positives, alignment length of at least 50% of the longer protein, and an e-value more significant than the most significant non-Hymenopteran hit to the same protein. To identify the members of a given paralogous family, we use a graph-based algorithm implemented in Perl using the Graph module, as follows: first, we convert blast hits to a graph, by creating edges between any pair of genes that are reciprocally connected by blast hits as defined above. Next, we extract the connected components of this overall graph. For each connected component with more than 2 members, we search for edges (which represent blastp hits) that if removed would increase the number of connected components, and remove the one with the lowest bitscore until either no more bridges exist (bridges are edges in a graph that if removed would increase the number of connected components) or all subgraphs are of size 1 or 2. We then define a gene family as the members of each subgraph.

**Signal peptide presence.** We define the presence of a signal peptide for each gene in the *Nasonia* proteome using signalp with the following options: -f short -s notm -t euk -c 70 -M 10. We analyze all isoforms of each protein, and consider a gene as having a signal peptide if any isoform has evidence for a signal peptide.

**Characterizing the properties of proteins encoded by highly induced genes.** To understand the properties of highly induced genes encoding short proteins, we used several tools designed to characterize different protein properties. We computed net charge and hydrophobicity based on the results of the AMP database server (http://aps.unmc.edu/AP/main.php); we computed disulfide bond formation with DISULFIND (http://disulfind.dsi.unifi.it/), and we estimated protein secondary structure with the default consensus prediction from the NPS server (http://npsa-pbil.ibcp.fr/cgi-bin/npsa_automat.pl?page = /NPSA/npsa_seccons.html).

## Supporting Information

**Table S1** Homology-based annotation of immune genes in *Nasonia*.
(XLS)

**Table S2** Genes differentially regulated by infection in *Nasonia*.
(XLS)

**Table S3** Properties of potential novel AMPs.
(XLS)

**Table S4** Data sources for blastp searches.
(XLS)

**Dataset S1** Full dataset.
(GZ)

## Acknowledgments

The Illumina library preparation and sequencing was done by the Cornell Biotechnology Resource Center. Read mapping, blast searches, and other computationally intensive algorithms were run on the Odyssey cluster supported by the FAS Science Division Research Computing Group at Harvard University. We would like to acknowledge Chris Childers and NasoniaBase for data hosting and community support.

## Author Contributions

Conceived and designed the experiments: TBS JHW AGC. Performed the experiments: TBS JHW. Analyzed the data: TBS. Contributed reagents/materials/analysis tools: AGC. Wrote the paper: TBS. Edited the manuscript and approved the final draft: TBS JHW AGC.